\documentclass{article}
\usepackage{spconf,amsmath,graphicx}

\usepackage{cite}
\usepackage{amsmath,amssymb,amsfonts}
\usepackage{url}
\usepackage{algorithmic}
\usepackage{textcomp}
\usepackage{xcolor}
\usepackage{tcolorbox}
\usepackage{booktabs} % To thicken table lines
\usepackage{amsthm} % Required for the proof environment
\usepackage{makecell, cellspace}
\usepackage{caption}
\usepackage{physics}
\usepackage{multirow}
\usepackage{hyperref} % 
\usepackage{cleveref}
\usepackage{adjustbox}
\usepackage{array}
\usepackage{subcaption}
\title{Dynamic resolution switching for live streaming}
\name{Xin Xiong, Yixu Chen, Hai Wei, Yongjun Wu, Sriram Sethuraman 
\thanks{\textcopyright\ 2026 IEEE. Personal use of this material is permitted. Permission from IEEE must be obtained for all other uses. Accepted to the 2026 IEEE International Conference on Image Processing (ICIP). Xin's work was done during an internship at Amazon Prime Video.}}
\address{% $^{1}$University of Southern California, Los Angeles, USA\\
         Amazon Prime Video, Seattle, USA}
\begin{document}
\ninept
%\ninept
%
\maketitle
\begin{abstract}
Conventional adaptive bitrate (ABR) streaming systems typically rely on static bitrate ladders to optimize Quality of Experience (QoE). 
While operationally simple, this ``one-size-fits-all'' approach neglects content-specific characteristics, often compromising streaming efficiency.
Per-title optimization methods address this by predicting the rate-distortion convex hull directly from the source content, but their reliance on pre-encoding source analysis can limit their applicability to live streaming.
Moreover, the objective video quality metrics (VQMs) they rely on are optimized for overall correlation with subjective scores rather than cross-over accuracy, often yielding inaccurate cross-over predictions and suboptimal ladder construction.
% Moreover, they rely on objective video quality metrics (VQMs) that are optimized for overall correlation with subjective scores rather than for cross-over accuracy.
% To address this, per-title optimization methods have emerged, attempting to predict the rate-distortion convex hull directly from source content. However, a limitation of these approaches is their fundamental reliance on objective video quality metric (VQM) as a proxy for ground truth. 
% Due to an inherent bias toward higher resolutions, these VQMs often yield inaccurate predictions of subjective resolution cross-over points, resulting in suboptimal ladder construction.
To overcome both limitations, we introduce a Dynamic Resolution Switching (DRS) framework for live streaming that remains fully compatible with existing streaming protocols. 
% Our approach constructs dynamic ladders by augmenting static ladders with strategically selected additional representations guided by user bandwidth distributions and cross-over regions, and analyzing the quality of the representations in real time. 
Our approach augments static ladders with strategically selected representations guided by user bandwidth distributions and cross-over regions. The quality of these representations is then analyzed in real time to construct dynamic ladders.
Central to this framework is a lightweight, bitstream-based VQM that ensures computational efficiency while maximizing the accuracy of subjective resolution cross-over prediction through training on Pairwise Comparison (PC) datasets.
At each bitrate, the VQM evaluates all candidate representations to identify the resolution maximizing the quality score. This decision process, operating at a configurable granularity (e.g., per segment), drives the dynamic resolution switching mechanism specifically optimized for the metric. Experimental results validate the approach, demonstrating a significant performance gain (approximately 9\% Bjøntegaard Delta rate reduction under the proposed VQM) while maintaining practical feasibility for live streaming.
\end{abstract}

\section{Introduction}
\label{sec:intro}

Video streaming has become an integral part of our daily lives. Due to the varying network bandwidth and device types among end-users, service providers widely implement adaptive bitrate (ABR) methods \cite{bentaleb2018survey} to enhance the quality of experience (QoE). The cornerstone of an ABR implementation is the bitrate ladder, a set of bitrate-resolution pairs (representations) selected dynamically to accommodate varying bandwidth.
The conventional approach to bitrate ladder construction involves the implementation of a static or ``one-size-fits-all'' ladder, utilizing predetermined bitrate-resolution pairs across all content. However, this methodology fails to account for content-specific characteristics, potentially yielding suboptimal results. To address this limitation, per-title optimized bitrate ladder construction has been proposed \cite{de2016complexity,menon2023just,katsavounidis2021iterative}. 
These methods exhaustively search the rate-distortion (RD) convex hull across pre-encoded copies to maximize QoE.
%These methods entail generating multiple pre-encoded copies and conducting an exhaustive search for the convex hull of rate–distortion (RD) operating points that maximizes QoE. 
However, the computational demands of these approaches make them impractical for live streaming applications.

\begin{figure*}[thbp]
    \centering
    \includegraphics[width=1\linewidth]{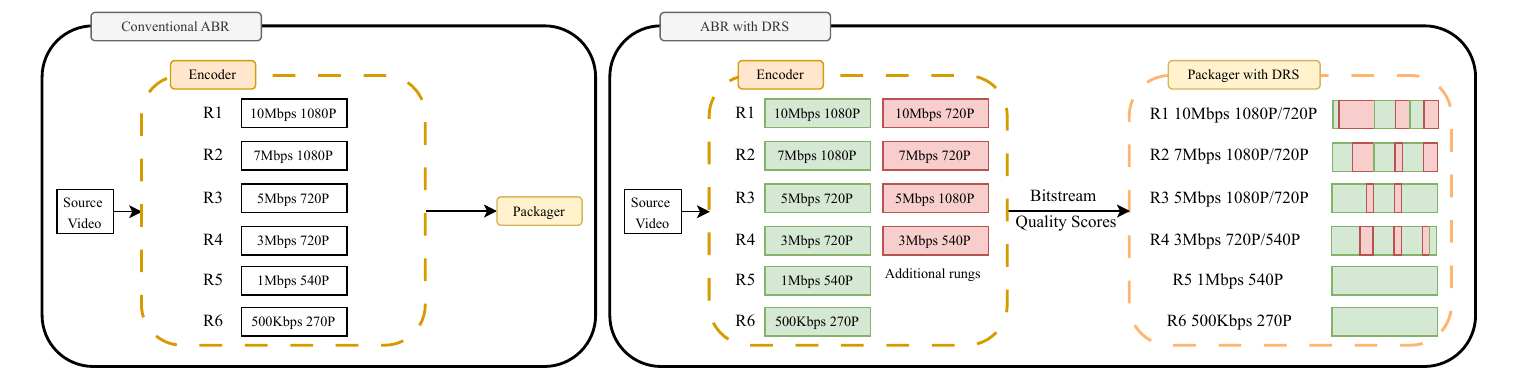}
    \caption{The proposed dynamic resolution switching pipeline in live streaming. Compared to conventional ABR systems, our approach introduces additional encoded representations and utilizes a bitstream-based VQM to predict quality scores, which are then used for adaptive resolution selection at a configurable granularity (e.g., per segment) to maximize QoE.
    % Link to the figure https://drive.google.com/file/d/1x-nLGYptP-I5b6Btbuk8vb5y7PG2E3SN/view?usp=sharing
    }
    \label{fig:enter-label}
    \vspace{-5mm}
\end{figure*}

Alternatively, some prior work has proposed to predict the convex hull directly from source content for ladder construction \cite{telili2025convex, yang2024optimal}. For instance, \cite{telili2025convex} estimates resolution cross-over points, defined as the intersections of per-resolution RD curves, from content features in order to maximize QoE. 
While these approaches reduce the computational burden, they still require pre-encoding source analysis, limiting their applicability to live streaming.
% While these approaches effectively reduce the computational burden of brute-force searching, they require source-side analysis prior to encoding, limiting their applicability to live streaming. 
Moreover, they typically rely on objective VQMs (e.g., VMAF \cite{vmaf2016} and SSIM \cite{wang2004image}) as proxies for ground-truth, primarily due to the labor-intensive nature of subjective studies.
%While these methods reduce the computational burden of brute-force searching, the required pre-processing (including feature extraction, model inference, and pre-encoding probes) introduces latency and complexity that preclude their application in live streaming scenarios. Furthermore, given the labor-intensive nature of subjective studies, such methods typically rely on objective VQMs, like VMAF, as a proxy for ground truth. 
However, recent work indicates that VQMs, including VMAF and P1204.3 \cite{rao2020bitstream}, often fail to accurately predict resolution cross-over points due to an inherent bias toward higher resolutions \cite{chen2024encoder}. Additionally, conventional VQMs are trained on Absolute Category Rating (ACR) datasets. 
% Crucially, research suggests that crossover points derived from ACR scores are suboptimal \cite{zhu2025video}. 
% In contrast, robust approaches utilizing Pair Comparison (PC) with active sampling have generated datasets that demonstrate a stronger correlation with subjective crossover points.
% and active sampling and cross-over customized ladder has yielded a dataset demonstrating strong correlation with observed subjective cross-over points \cite{zhu2025video}.
% Crucially, recent findings indicate that cross-over points derived from ACR scores are suboptimal \cite{zhu2025video}, whereas datasets constructed via pairwise comparison (PC) with active sampling achieves better cross-over accuracy.
% exhibit a much stronger correlation with subjective resolution cross-over ground truth.
Cross-over points derived from ACR scores are suboptimal \cite{zhu2025video}; pairwise comparison (PC) datasets achieve better cross-over accuracy.
Another common direction for content-adaptive encoding is variable bitrate (VBR) encoding \cite{netflix2026vbr}, which saves bits during low-complexity scenes. However, VBR primarily optimizes bit allocation within a single representation rather than across representations, leaving the resolution selection problem unaddressed. Furthermore, while VBR reduces average bandwidth, peak CDN capacity must still be provisioned for the worst-case content complexity.

While prior cross-over-focused work \cite{zhu2025video, chen2024encoder} has been primarily diagnostic, none has closed the loop from cross-over analysis to an operational real-time system. To bridge this gap, we propose a dynamic resolution switching (DRS) framework tailored for live streaming, transitioning cross-over insights from diagnosis to action. Furthermore, DRS achieves end-to-end bit savings from origin to edge to client and is complementary to both constant bitrate (CBR) and VBR encoding.
% To address these limitations, in this work, we propose a dynamic resolution switching (DRS) framework tailored for live streaming applications. Unlike approaches requiring a variable number of representations per segment, our method maintains compatibility with existing streaming protocols. 
% The framework focuses on optimizing resolution selection within critical crossover bitrate regions.
\autoref{fig:enter-label} illustrates the proposed DRS pipeline integrated within the ABR streaming system. 
In contrast to conventional ABR approaches, we augment the standard static ladder with a data-driven set of additional representations, to better utilize the unused capacity of the encoding server, especially for low-complexity codecs such as AVC \cite{wiegand_overview_2003}. 
During encoding, the encoder generates an expanded set of representations, each embedded with a computed quality score. The packager then dynamically filters these streams, selecting the resolution with the highest quality score for each bitrate. 
This filtering reduces the ladder size back to its standard static dimensions before delivery, thereby making our method compatible with existing live streaming protocols.
The efficacy of this quality-driven resolution selection mechanism hinges on a VQM that balances computational efficiency with high accuracy in predicting subjective resolution cross-over points. 
% Pixel domain full-reference metrics, such as VMAF and SSIM, typically require the signal to be fully decoded and rescaled to the source resolution. This process introduces additional computational overhead, which can be prohibitive for live streaming applications. Conversely, 
Although standardized bitstream-based metrics like P1204.3 are computationally efficient, they often lack the accuracy required to reliably predict subjective resolution cross-over points.
% Pixel domain full reference based metrics, such as VMAF, SSIM, usually require fully decoded signal, and the rescale to the source resolution for computation, which may add up computation complexity during encoding in live streaming. Bit-stream based metrics such as P1204.3 is efficient but suffer from the performance at predicting subjective cross-over points.
To address this limitation, we adopt the Encoder-Quantization-Motion (EQM) model \cite{chen2024encoder} and recalibrate it on the PC dataset from \cite{zhu2025video} for improved cross-over accuracy. The primary contributions of this work are summarized as follows:
\begin{itemize}
    % \item We propose a DRS pipeline that augments static ladders with strategically selected additional representations. The pipeline uses AVC-EQM to dynamically identify and select the optimal resolution to maximize QoE for live streaming.
    % \item We develop AVC-EQM, an efficient bitstream-based VQM tailored for AVC streams. This metric has low computational complexity with minimal overhead while demonstrating a high correlation with subjective resolution cross-over points.
    \item A DRS pipeline that is compatible with existing streaming protocols, which augments static bitrate ladders with strategically selected additional representations and dynamically selects the optimal resolution per segment to maximize QoE.
    \item AVC-EQM, an efficient bitstream-based VQM tailored for the DRS pipeline, with minimal overhead and a high correlation with subjective resolution cross-over points.
    We also introduce new evaluation measures that align cross-over assessment with the operational objective of DRS.
    \item Experiments on real-world sports videos demonstrate that our proposed DRS pipeline significantly improves QoE. Compared to a baseline static ladder, we achieve approximately 9\% Bjøntegaard Delta rate (BD-rate) savings and a 0.15 gain in BD-AVC-EQM (on a 0--10 scale).
\end{itemize}

\section{AVC-EQM Model}
% \textbf{Ongoing work:} Adding extrapolation at lower bitrate range to reduce the edge cases where the no crossover is covered by the subjective ladder design.

% Todo: Should we have a table comparing the previous table vs improved table for the ranking change, and the number of edge cases (double crossover/no crossover) that's being reduced as a quantifiable metrics for this improvement?
\subsection{Feature selection}
\begin{figure}[!thbp]
    \centering
    \includegraphics[width=1\linewidth]{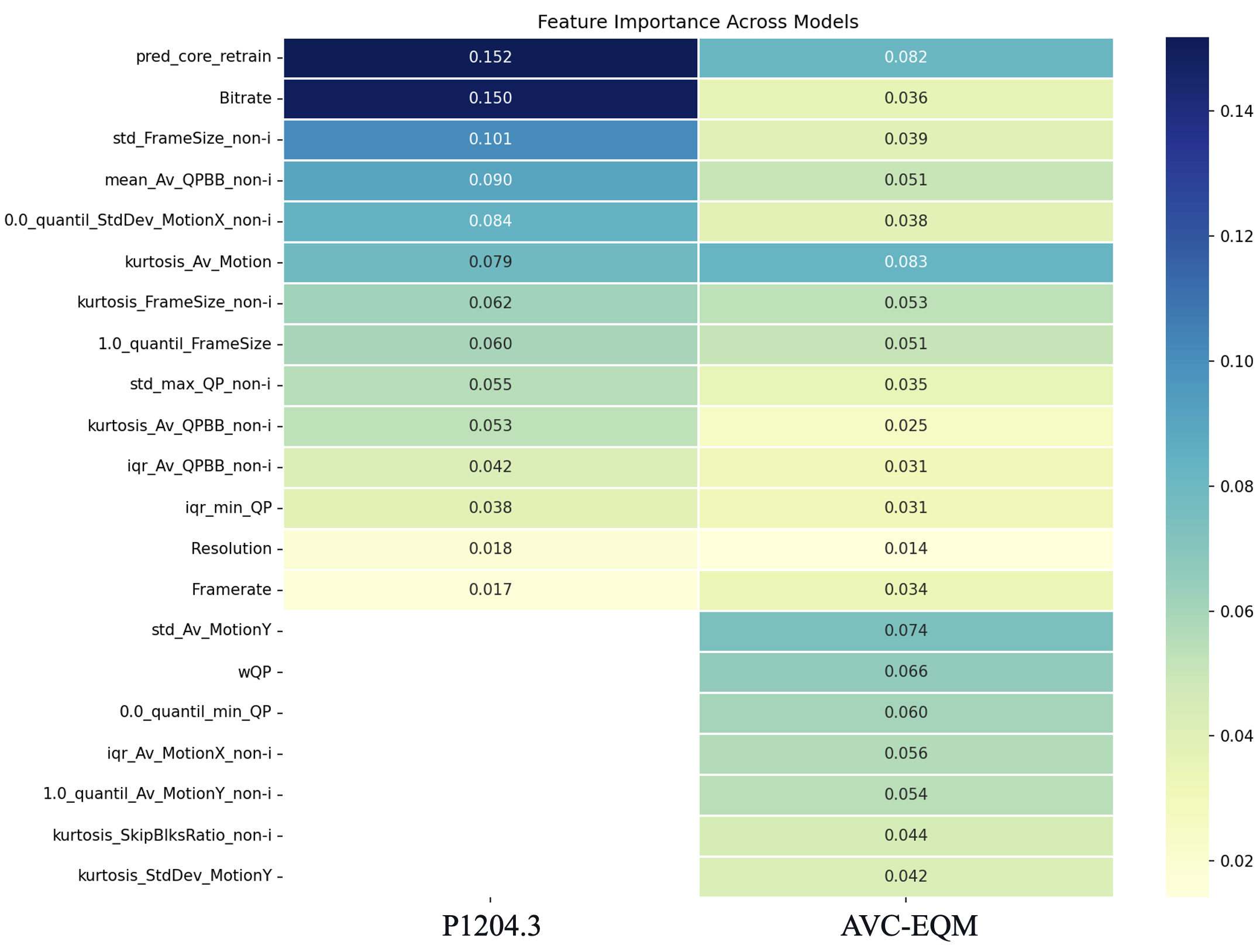}
    \caption{Feature importance comparison (random forest residual model): retrained P1204.3 vs. AVC-EQM on the LSCO-AVC dataset. AVC-EQM leverages a broader set of significant features.
    }
    \label{fig:importance}
    \vspace{-3mm}
\end{figure}

Although EQM \cite{chen2024encoder} was initially developed for HEVC \cite{sullivan_overview_2012}, the strict real-time constraints of live streaming often limit the feasibility of HEVC in high-density encoding scenarios. In contrast, AVC remains a more practical choice for our DRS pipeline due to its lower encoding complexity. In this work, we adapt the EQM model \cite{chen2024encoder} from HEVC to AVC, and propose a specialized AVC-EQM. Given the inherent differences between the two codecs, such as hierarchical block partitioning structures and motion vector prediction mechanisms, we first identify the subset of EQM features in \cite{chen2024encoder} that are applicable to AVC. 
We then employ Greedy Feature Selection (GFS) \cite{guyon2003introduction} on the Live Sport Cross-Over (LSCO) dataset \cite{zhu2025video}, whose quality labels are generated via PC to ensure a better correlation with subjective resolution cross-over, in order to determine the optimal feature set for the AVC bitstream.
% Subsequently, utilizing the Live Sport Cross-Over (LSCO) dataset \cite{zhu2025video}, whose quality labels are generated via PC to ensure better correlation with subjective resolution cross-over, we employ Greedy Feature Selection (GFS) \cite{guyon2003introduction} to determine the optimal feature set for AVC bitstream. 
The selected features are shown in \autoref{fig:importance}. The features that are exclusive to AVC-EQM exhibit significant importance, suggesting that they were likely overlooked by P1204.3, since its feature set was neither designed nor trained on datasets specifically targeting subjective resolution cross-over.
To validate the effectiveness of AVC-EQM, we evaluate it from two perspectives: (i) its correlation with subjective resolution cross-over points, and (ii) its runtime complexity.

\begin{table*}[t]
\caption{RCQL benchmark results on the LSCO-AVC dataset. P1204.3 and AVC-EQM are bitstream-based metrics, whereas the other VQMs are computed by upscaling decoded frames to the source resolution for full-referenced quality evaluation. }
\label{tab:RCQL_AVC_Crossover}
% \vspace{-0.2cm}
\begin{adjustbox}{width=1.8\columnwidth,center}
\begin{tabular}{@{}l|llllllll|ll@{}}
\toprule
& \multicolumn{8}{c|}{Out-of-box} & \multicolumn{2}{c}{Retrained} \\ \cmidrule(l){2-9} \cmidrule(l){10-11}

 &
   &
  PSNRy &
  SSIM &
  MS-SSIM &
  VMAF &
  VMAF 4k &
  \begin{tabular}[c]{@{}l@{}}VMAF\\ 4k neg\end{tabular} & P1204.3 & P1204.3* & AVC-EQM \\ \midrule

\multirow{2}{*}{Correlation$\uparrow$}
& SROCC & 0.728 & 0.815 & 0.812 & 0.826 & 0.822 & 0.824 & 0.895 & 0.912 & \textbf{0.926} \\
& PLCC & 0.684 & 0.741 & 0.781 & 0.832 & 0.825 & 0.827	& 0.895 & 0.906 & \textbf{0.925} \\ \midrule
\multirow{2}{*}{$\Delta$Bitrate (Kbps)$\downarrow$}
& 1080p vs 720p&  \textbf{862.66}& 8153.23& 4078.12& 1193.89& 1131.49& 1117.85& 1512.71 & 1404.45 & 1195.27 \\
& 720p vs 540p&  \textbf{254.9} & 4191.05& 783.58& 377.51& 371.39& 366.8& 644.11 & 465.1 & 484.64\\\midrule
\multirow{2}{*}{$RCQL_{s}$ (JOD$\times$Kbps)$\downarrow$} 
& 1080p vs 720p&  \textbf{178.49}& 4825.25& 1701.84& 293.41& 275.68& 267.03& 758.38 & 458.98 & 402.02\\
& 720p vs 540p&  \textbf{51.33} & 3900.03& 349.07& 150.92& 136.34& 135.73& 336.31 & 158.97 & 171.67\\\midrule
\multirow{2}{*}{$RCQL_{avg}$ (JOD)$\downarrow$} 
& 1080p vs 720p&  \textbf{0.1674}& 0.5694& 0.3299& 0.1988& 0.2022& 0.2006& 0.254 & 0.2854 & 0.2722 \\
& 720p vs 540p&  \textbf{0.1564}& 0.7573& 0.3103& 0.2037& 0.188& 0.1871& 0.3493 & 0.2695 & 0.2873\\\midrule
% & 1080p vs 720p& \textbf{903.33}& 8111.57& 4039.23& 1234.56& 1172.17& 1158.52& 1472.56 & 1444.11 & 1159.18\\
% & 720p vs 540p& \textbf{254.9}& 4191.05& 783.58& 377.51& 371.39& 366.8& 644.11 & 465.10 & 484.64 \\\midrule
% & 1080p vs 720p& \textbf{202.67}& 4793.0& 1687.0& 317.58& 299.86& 291.21& 739.67 & 483.15 & 377.10\\
% & 720p vs 540p& \textbf{51.33} & 3900.03& 349.07& 150.92& 136.34& 135.73& 336.31 & 158.97 & 171.67\\\midrule
% & 1080p vs 720p& \textbf{0.1749}& 0.5671& 0.3203& 0.2062& 0.2097& 0.208& 0.2448 & 0.2929 & 0.2668\\
% & 720p vs 540p& \textbf{0.1564}& 0.7573& 0.3103& 0.2037& 0.188& 0.1871& 0.3493 & 0.2695 & 0.2873\\\midrule
\multirow{2}{*}{$Acc$ (\%)$\uparrow$} 
& 1080p vs 720p& 93.55& 58.06& 74.19& 91.94& \textbf{95.16}& \textbf{95.16}& 88.71 &  89.17 & 92.08 \\
& 720p vs 540p& \textbf{86.36} & 62.12& 80.3& 83.33& \textbf{86.36}& \textbf{86.36}& 80.3 &  84.17 & 79.17\\\bottomrule
\multirow{2}{*}{$QL$ (JOD)$\downarrow$} 
& 1080p vs 720p& 0.214 & 0.383& 0.361& 0.658& 0.768& 0.768& 0.287 & 0.172 & \textbf{0.085} \\
& 720p vs 540p& 0.299& 0.546& 0.305& 0.374& 0.272 & 0.272& 0.39 & \textbf{0.159} & 0.217\\\bottomrule

% \multirow{2}{*}{$\Delta$Bitrate (Kbps)$\downarrow$}
% & 1080p vs 720p& 930.39& 2175.86& 1397.96& 1143.09& \textbf{760.83}& 804.18& 1130.32\\
% & 720p vs 540p& \textbf{254.9}& 2975.96& 1252.78& 347.16& 362.18& 359.54& 673.25\\\midrule
% \multirow{3}{*}{$RCQL_{s}$ (JOD$\times$Kbps)$\downarrow$} 
% & 1080p vs 720p& 195.89& 1116.03& 435.55& 281.97& \textbf{154.52}& 165.22& 357.26\\
% & 720p vs 540p& \textbf{51.33}& 2308.83& 663.61& 130.28& 131.34& 129.55& 382.28\\\midrule
% \multirow{3}{*}{$RCQL_{avg}$ (JOD)$\downarrow$} 
% & 1080p vs 720p& 0.1715& 0.4062& 0.2504& 0.2018& \textbf{0.1574}& 0.1631& 0.2393\\
% & 720p vs 540p& \textbf{0.1564}& 0.5052& 0.2652& 0.1892& 0.1858& 0.1821& 0.34\\\bottomrule

\end{tabular}
\end{adjustbox}
\end{table*}

\begin{table*}[t]
\caption{RCQL benchmark results on the LSCO-HEVC dataset. EQM NR and EQM FR are adopted from \cite{chen2024encoder}. HEVC-EQM denotes the EQM NR model retrained on the LSCO-HEVC dataset.}
\label{tab:RCQL_HEVC_Crossover}
% \vspace{-0.2cm}
\begin{adjustbox}{width=1.8\columnwidth,center}
\begin{tabular}{@{}l|llllllllll|ll@{}}
\toprule
& \multicolumn{10}{c|}{Out-of-box} & \multicolumn{2}{c}{Retrained} \\ \cmidrule(l){2-11} \cmidrule(l){12-13}

 &
   &
  PSNRy &
  SSIM &
  MS-SSIM &
  VMAF &
  VMAF 4k &
  \begin{tabular}[c]{@{}l@{}}VMAF\\ 4k neg\end{tabular} &
  P1204.3 &
  EQM NR &
  EQM FR & P1204.3* & HEVC-EQM\\ \midrule

\multirow{2}{*}{Correlation$\uparrow$}
& SROCC & 0.705 & 0.820	& 0.857 & 0.880	& 0.876 & 0.877 & 0.851 & 0.921 & 0.928 & 0.908 & \textbf{0.943}\\
& PLCC & 0.673 & 0.652 & 0.748 & 0.873 & 0.866 & 0.869 & 0.830	& 0.916 & 0.922 & 0.901 & \textbf{0.943} \\ \midrule
\multirow{2}{*}{$\Delta$Bitrate (Kbps)$\downarrow$}
& 1080p vs 720p& \textbf{1277.47}& 5479.27& 2433.18& 1515.0& 1511.07& 1377.85& 1488.57& 1595.54& 2171.01 & 1519.58 & 1484.72\\
& 720p vs 540p& 560.69& 2295.09& 755.37& 595.62& 644.27& 627.74& 716.18& 584.58& \textbf{544.54} & 875.86 & 811.07\\\midrule
\multirow{2}{*}{$RCQL_{s}$ (JOD$\times$Kbps)$\downarrow$} 
& 1080p vs 720p& 560.32 & 3581.81& 1401.83& 615.34& 598.65& 578.82& 645.7& 767.43& 1440.03 & \textbf{477.44} & 556.65\\
& 720p vs 540p& \textbf{112.27}& 1843.95& 255.25& 115.58& 131.22& 123.82& 184.58& 131.76& 120.34 & 312.88 & 216.91\\\midrule
\multirow{2}{*}{$RCQL_{avg}$ (JOD)$\downarrow$} 
& 1080p vs 720p& 0.2549& 0.5956& 0.3173& 0.2596& 0.2587& 0.2515& \textbf{0.2182}& 0.309& 0.3858 & 0.2571 & 0.2508\\
& 720p vs 540p& 0.1193& 0.4875& 0.1942& 0.0918& 0.1039& \textbf{0.1012}& 0.1995& 0.1656& 0.1441 & 0.2696 & 0.2646\\ \midrule
\multirow{2}{*}{$Acc$ (\%)$\uparrow$} 
& 1080p vs 720p& \textbf{87.5}& 60.94& 79.69& 84.38& 85.94& 85.94& 85.94& 79.69& 81.25 & 78.33 & 80.83 \\
& 720p vs 540p& 80.3& 68.18& 80.3& \textbf{84.85}& \textbf{84.85}& \textbf{84.85}& 78.79& 75.76& 81.82 & 76.25 & 76.25 \\\bottomrule
\multirow{2}{*}{$QL$ (JOD)$\downarrow$} 
& 1080p vs 720p& 0.249 & 0.517& 0.29& 0.294& 0.321& 0.321& 0.39& 0.341& 0.365 & 0.18 & \textbf{0.164} \\
& 720p vs 540p& 0.226 & 0.478& 0.241& 0.23& 0.23& 0.23& 0.29& 0.347& 0.345 & \textbf{0.176} & 0.200 \\\bottomrule

\end{tabular}
\end{adjustbox}
\end{table*}

\subsection{Benchmark on cross-over dataset}

To assess the correlation between objective VQMs and subjective cross-over points, Zhu \emph{et al.} \cite{zhu2025video} introduced the Resolution Cross-over Quality Loss (RCQL) to quantify the quality loss caused by cross-over prediction errors. RCQL employs the Piecewise Cubic Hermite Interpolating Polynomial (PCHIP) \cite{fritsch1984method} to interpolate the Rate-Distortion (RD) curves, identifying ground-truth subjective cross-over points at the curve intersections. While PCHIP is standard for calculating the BD-rate, it has limitations when handling noisy subjective quality labels recovered by Bradley-Terry or PWCMP models \cite{perez2017practical}, which often exhibit non-monotonic behavior with respect to bitrate \cite{brand2023interpolation}. In such scenarios, PCHIP often yields unreliable results, potentially producing multiple ambiguous intersections because of the violation of monotonicity.
% \textbf{The left subfigure illustrates multiple intersections in the interpolated Rate-Distortion (RD) curves, resulting in triple cross-over points}. While the implementation in \cite{zhu2025video} adopts the intersection point having the highest rate, yielding a cross-over bitrate of approximately 12.5 Mbps, this threshold appears substantially higher than typical expectations for the resolution cross-over between 1080p and 720p.
% Moreover, the cross-over quality loss—depicted by the pink-shaded regions in Figure \ref{fig:Exp_curve_1}—serves as a metric for evaluating cross-over prediction performance. The left subfigure reveals two distinct, disconnected regions that, despite their geometric similarity, represent fundamentally different perceptual phenomena. This bifurcation not only complicates the interpretation of results but also introduces additional computational challenges in the assessment framework.

To address this limitation, we adopt a more robust alternative using a 4-parameter logistic function for fitting the ground-truth RD curve:
\begin{equation}
f(x) = \beta_2 + \frac{\beta_1 - \beta_2}{1 + \exp\left(-\frac{x-\beta_3}{|\beta_4|}\right)}
\end{equation}
where $\beta_3$ represents the horizontal inflection point. To ensure that the fitted curve models the saturation characteristic of RD curves, we constrain the optimization domain of $\beta_3$ to the interval $[r_{min}/2, r_{min}]$, where $r_{min}$ denotes the minimum bitrate present in each RD curve. 
Together with the inherent monotonicity of the logistic function, this enables more accurate cross-over determinations.
% This procedure ensures monotonicity of the fitted subjective RD curves, and thereby enables more accurate cross-over determinations.

Leveraging the refined subjective cross-over points, we re-evaluated the performance of various VQMs. The results are summarized in \autoref{tab:RCQL_AVC_Crossover} and \autoref{tab:RCQL_HEVC_Crossover}.
Following the RCQL evaluation in \cite{zhu2025video}, three key metrics are reported:
(1) $\Delta$Bitrate (Kbps)$\downarrow$ quantifies the absolute bitrate deviation between the subjective cross-over (determined from the fitted subjective RD curves) and the objective cross-over predicted by the VQM;
(2) $RCQL_{s}$ (JOD$\times$Kbps)$\downarrow$ represents the cumulative quality loss, calculated as the integral area bounded by the two RD curves within the bitrate interval defined by the subjective and objective cross-over points; and
(3) $RCQL_{avg}$ (JOD)$\downarrow$ denotes the mean quality loss, calculated as the average perceptual loss in the cross-over divergence region.

While the RCQL metric introduced in \cite{zhu2025video} offers valuable insights into a VQM's ability to locate exact resolution cross-over points, the primary objective of our DRS framework is slightly different. Rather than pinpointing the precise cross-over bitrate, our focus is on accurately identifying the optimal resolution at any given operating bitrate. To better align with this objective, we propose a complementary RCQL metric that quantifies the accuracy of relative quality ranking across resolutions. This methodology involves pairwise comparisons of quality scores between different resolutions (e.g., 720p vs. 1080p) at the same bitrate. For each pair, we evaluate the concordance between the objective VQM's preference and the ground-truth subjective quality ranking. The resulting accuracy is denoted as $Acc$ (\%).
In instances where the objective VQM incorrectly predicts the perceptual ranking (e.g., favoring 1080p when subjective scores indicate that 720p is superior), we quantify the magnitude of the subjective quality difference between the two resolutions. We then define $QL$ (JOD)$\downarrow$ as the mean quality degradation averaged across all such incorrectly predicted events. The proposed complementary RCQL measures are particularly pertinent to the DRS framework, where decision-making relies directly on VQM scores. By quantifying the quality loss incurred by suboptimal resolution selection, they provide a practical evaluation aligned with operational DRS in live streaming. 

Table \ref{tab:RCQL_AVC_Crossover} presents the RCQL benchmark on the LSCO-AVC dataset. P1204.3* and AVC-EQM are trained and tested with 1000 runs of 5-fold cross-validation. The cross-validation procedure uses a content-based split, in which videos generated from the same source content are kept in the same group. For each RCQL metric, we first take the median (over the 1000 runs) for each source content and then average across all contents.
In terms of overall correlation, AVC-EQM achieves the highest SROCC and PLCC, outperforming the other VQMs. While P1204.3* shows a slight advantage on the lower-resolution cross-over (720p vs. 540p), AVC-EQM is  significantly better on the more critical high-resolution cross-over (1080p vs. 720p). Specifically, AVC-EQM reduces $QL$ by approximately 50\% relative to P1204.3* (0.085 vs. 0.172). This indicates that AVC-EQM is particularly well suited to our proposed DRS pipeline.
Furthermore, in light of the new subjective cross-over determination method and the new RCQL measures, we also re-benchmark the original HEVC bitstream-based EQM \cite{chen2024encoder} on the LSCO-HEVC dataset (\autoref{tab:RCQL_HEVC_Crossover}). Following a similar GFS process, the retrained HEVC-EQM generally surpasses the other bitstream-based metrics at identifying subjective cross-over points for 1080p vs. 720p.

% Table \ref{tab:RCQL_AVC_Crossover} shows the benchmark of RCQL on the LSCO-AVC dataset. Models denoted with an asterisk (*) are trained and tested for 1000 times 5-fold cross validation. The cross-validation procedure employed content-based split, where videos generated from the same content will be considered as a group. For these retrained models, the evaluation methodology first computed the median performance for each source content independently, followed by averaging these medians across all source videos to obtain the final performance measures. 
% As we can see, the AVC-EQM has the best correlation and has better RCQL for 1080P vs 720P than P1204.3. 
% As we introduced new subjective cross-over point determination method and new RCQL measurement, we also re-benchmark the EQM for HEVC proposed in \cite{chen2024encoder} on the LSCO-HEVC dataset. Table \ref{tab:RCQL_HEVC_Crossover} shows the benchmark of cross-over accuracy on the LSCO-HEVC dataset. Similar to AVC-EQM retrain, a greedy feature selection (GFS) is performed to find better features for HEVC-EQM on LSCO-HEVC dataset. In genral, the HEVC-EQM* achieves better subjective cross-over performance than the other bitstream-based metrics.

\subsection{Runtime complexity of AVC-EQM}
AVC-EQM can operate on both the encoder side (facilitating DRS in live streaming) and the decoder side (serving as a research prototype). We implement the decoder side AVC-EQM in libavcodec and the encoder side AVC-EQM in x264. For efficiency, the decoder side implementation supports frame-level multi-threading. \autoref{tab:Speed_Comparison} presents a speed comparison between P1204.3 and AVC-EQM (using 15 threads) for AVC bitstreams on an AWS EC2 c7a.48xlarge instance. Results are averaged over 15 source videos (60\,fps, 8 seconds duration). Notably, for 1080p content, AVC-EQM is over $10\times$ faster than P1204.3. For the evaluation of encoder side integration, we use an 8-second 1080p SDR source video from the sports dataset on the same EC2 instance, with the x264 single-thread medium preset. As shown in Table \ref{tab:Speed_Comparison_Encoder}, the additional computational overhead of AVC-EQM is small, approximately 2\% to 4\%, which validates its suitability for our proposed DRS pipeline.

\begin{table}[!t]
\centering
\caption{Decoder side speed comparison on AVC bitstream. The proposed AVC-EQM is over 10 times faster than P1204.3.}
\setlength{\tabcolsep}{3pt}
\label{tab:Speed_Comparison}
\begin{tabular}{cccc}
\toprule
Resolution & Bitrate(Mbps) & P1204.3(FPS) & EQM(FPS) \\ \midrule
% \multirow{2}{*}{1080P}  & 9  & 74.414 & 6.975 \\
%  & 6  & 73.635 & 6.558 \\ \midrule
% \multirow{2}{*}{720P}  & 2.2  & 53.383 & 4.842 \\
%  & 1.6  & 53.323 & 4.731 \\ \midrule
% \multirow{2}{*}{540P}  & 1.2  & 46.596 & 4.292 \\
% & 0.6  & 46.529 & 4.172 \\
% \multirow{2}{*}{1080P}  & 9  & 6.45 & 68.82 \\
%  & 6  & 6.52 & 73.19 \\ 
% \multirow{2}{*}{720P}  & 2.2  & 8.99 & 99.13 \\
%  & 1.6  & 9.00 & 101.46 \\ 
% \multirow{2}{*}{540P}  & 1.2  & 10.30 & 111.84 \\
% & 0.6  & 10.32 & 115.05 \\
1080P  & 9  & 6.45 & 68.82 \\
1080P & 6  & 6.52 & 73.19 \\ \midrule
720P & 2.2  & 8.99 & 99.13 \\
720P & 1.6  & 9.00 & 101.46 \\ \midrule
540P & 1.2  & 10.30 & 111.84 \\
540P & 0.6  & 10.32 & 115.05 \\
\bottomrule
\end{tabular}
\end{table}

\begin{table}[!t]
\centering 
\caption{Encoder side runtime complexity analysis. AVC-EQM introduces a small overhead of only 2\% to 4\%, making it suitable for live streaming applications.}
\label{tab:Speed_Comparison_Encoder}
\begin{tabular}{ccccc}
\toprule
Res. & Bit. & x264(s) & x264\&EQM(s) & Overhead(\%) \\ \midrule
1080P  & 9  & 166.8 & 169.8 & 1.8 \\
720P  & 1.6  & 74.3 & 76.8 & 3.3 \\
540P & 0.6 & 40.8 & 42.0 & 2.9 \\
\bottomrule
\end{tabular}
\vspace{-3mm}
\end{table}

\section{Empirical evaluation}
As the proposed DRS pipeline augments a static ladder with a small number of additional representations, we first describe how these representations are selected in \autoref{sec:glsa}. We then evaluate the resulting performance in \autoref{prototype}.

\subsection{GOP-level statistical analysis }
\label{sec:glsa}
\begin{figure}[!t]
\centering
\includegraphics[width=8.5cm]{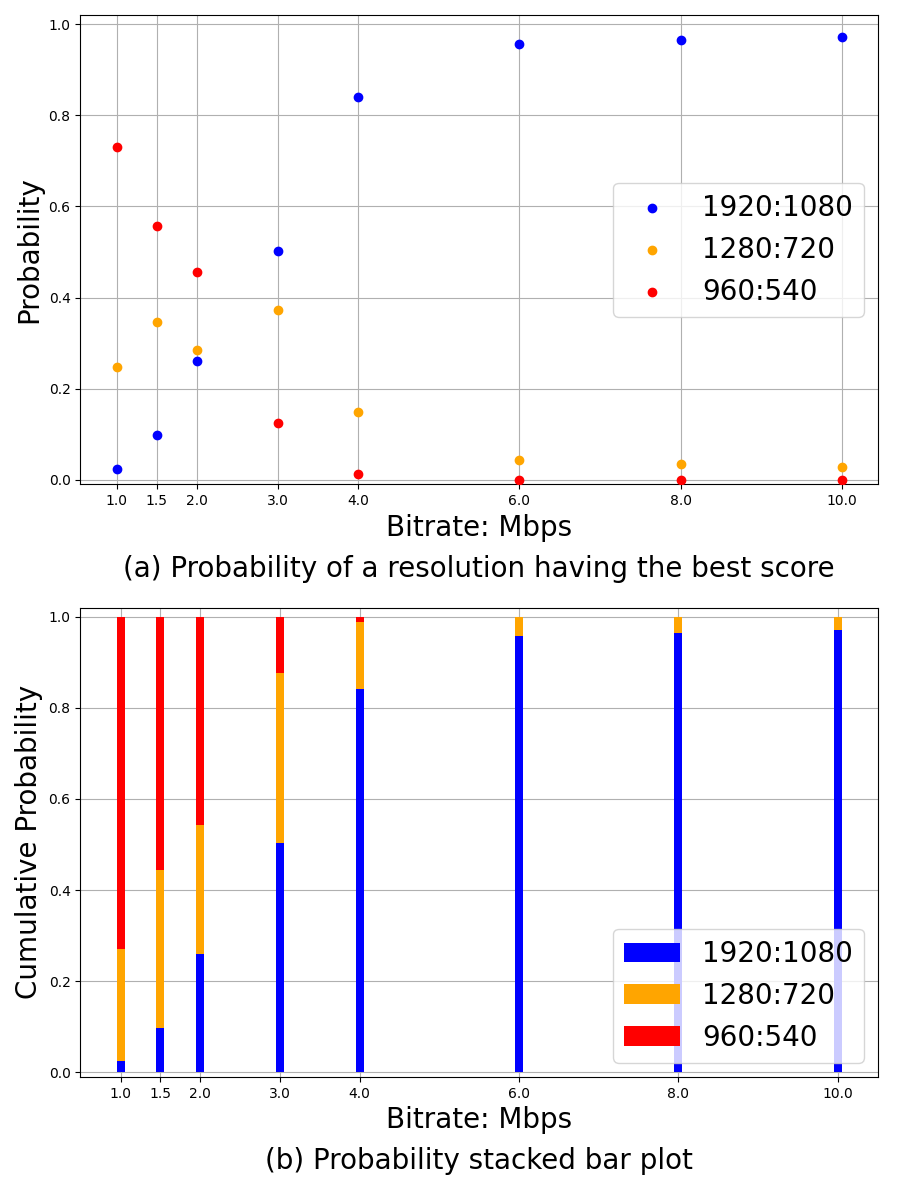}
\vspace{-3mm}
\caption{Statistical analysis of the resolution-quality relationship using AVC-EQM scores over 1800 GOPs of sports content. (a) Probability that each resolution yields the highest quality at a given bitrate. (b) Cumulative probability derived from (a), with resolutions distinguished by color.
}
\label{fig:TNF_EQM_Prob}
\vspace{-3mm}
\end{figure}
This analysis uses a fixed Group of Pictures (GOP) size of 1 second, consistent with standard live streaming protocols. A 30-minute video sequence is encoded using an AVC codec across a diverse set of representations. Focusing on bitrates within the typical resolution switching range (see the \textit{Rate} column in \autoref{tab:TNF_ladder}), the video is encoded at 540p, 720p, and 1080p. For each bitrate, we count the number of GOPs in which a specific resolution yields the highest quality as measured by AVC-EQM.

% \autoref{fig:TNF_EQM_Prob} (a) presents the probability of a given resolution achieving the highest VQM score at a specific bitrate. 
\autoref{fig:TNF_EQM_Prob}(a) presents, for each bitrate, the probability that a given resolution achieves the highest VQM score. 
This probability is defined as the ratio of GOPs at which the resolution offers the best VQM score to the total number of GOPs (1800 for the 30-minute sequence). \autoref{fig:TNF_EQM_Prob}(b) shows the corresponding cumulative probabilities. Taking 1080p as an example, at lower bitrates, only a few GOPs exhibit optimal quality at this resolution; however, as the bitrate increases, nearly all GOPs achieve best quality at 1080p. These findings suggest the design of augmented bitrate ladders for DRS: selecting the resolutions with a higher probability of optimality at each bitrate maximizes overall QoE as measured by the VQM. Given the strict constraints of live streaming, which limit the number of representations in the ladder, it is crucial to evaluate the marginal quality gain of each additional representation when optimizing the selection.

The optimization for representation selection can be formulated as:
\begin{equation}
\begin{aligned}
    & \underset{\Theta}{\text{maximize}} & & \sum_{r \in R} W_r \sum_{i=1}^N Q(i, r, \theta_{i, r}), \; \theta_{i,r} \in \Theta \\
    & \text{subject to} & & |\Theta| \leq K
\end{aligned}
\label{eq:opt_ladder}
\end{equation}
where $R$ is the set of bitrates in the ladder, $N$ is the total number of GOPs, $\theta_{i,r}$ is the representation that offers the best score for the $i$-th GOP at bitrate $r$, and $Q(\cdot)$ denotes the quality as measured by the VQM. $W_r$ is the weight for the corresponding bitrate, which can be derived from the real-world network bandwidth distribution. $K$ is the maximum number of representations in a ladder. The proposed formulation enables the construction of an augmented ladder that extends the static ladder for the DRS framework while optimizing for QoE. 
% The Table \ref{tab:TNF_ladder} shows the optimized ladder for sports content using EQM scores. 
% It should be noted that, after encoding, only the representationss that offer the best quality scores are kept for each segment at each bitrate. Thus, the final number of rerepresentationss (after selection) will be the same as the baseline ladder. 

\begin{table}[!thbp]
\centering
\caption{Enhancement of the baseline ladder for sports content using AVC-EQM scores. The \textit{content optimized} ladder selects the resolution with the highest probability of achieving the best AVC-EQM score at each bitrate (see \autoref{fig:TNF_EQM_Prob}), while the \textit{dynamic optimized} ladder extends the baseline by adding 4 additional representations, which serve as representation candidates in the DRS pipeline and are optimized via \autoref{eq:opt_ladder}.
}
\label{tab:TNF_ladder}
\begin{tabular}{cccc}
\toprule
Rate & Baseline ladder & Content Opt. & Dynamic Opt. \\
\midrule
1   & 540P  & 540P & 540P \\
1.5 & 540P  & 540P & 540P \\ \midrule
2   & 720P  & 540P & 540P, 720P \\
3   & 720P  & 1080P & 720P, 1080P   \\
4   & 720P  & 1080P & 720P, 1080P \\ \midrule
6   & 1080P & 1080P & 720P, 1080P \\
8   & 1080P & 1080P & 1080P \\
10  & 1080P & 1080P & 1080P \\
\bottomrule
\end{tabular}
\vspace{-1em}
\end{table}

% This formulation enables the design of an augmented static ladder, building upon the baseline ladder structure. Note that the number of rerepresentations is reduced to be the same as previous amount while keep the best resolution at the same bitrate as previous ladder for each segment.

\begin{figure}[!t]
\centering
\includegraphics[width=8.5cm]{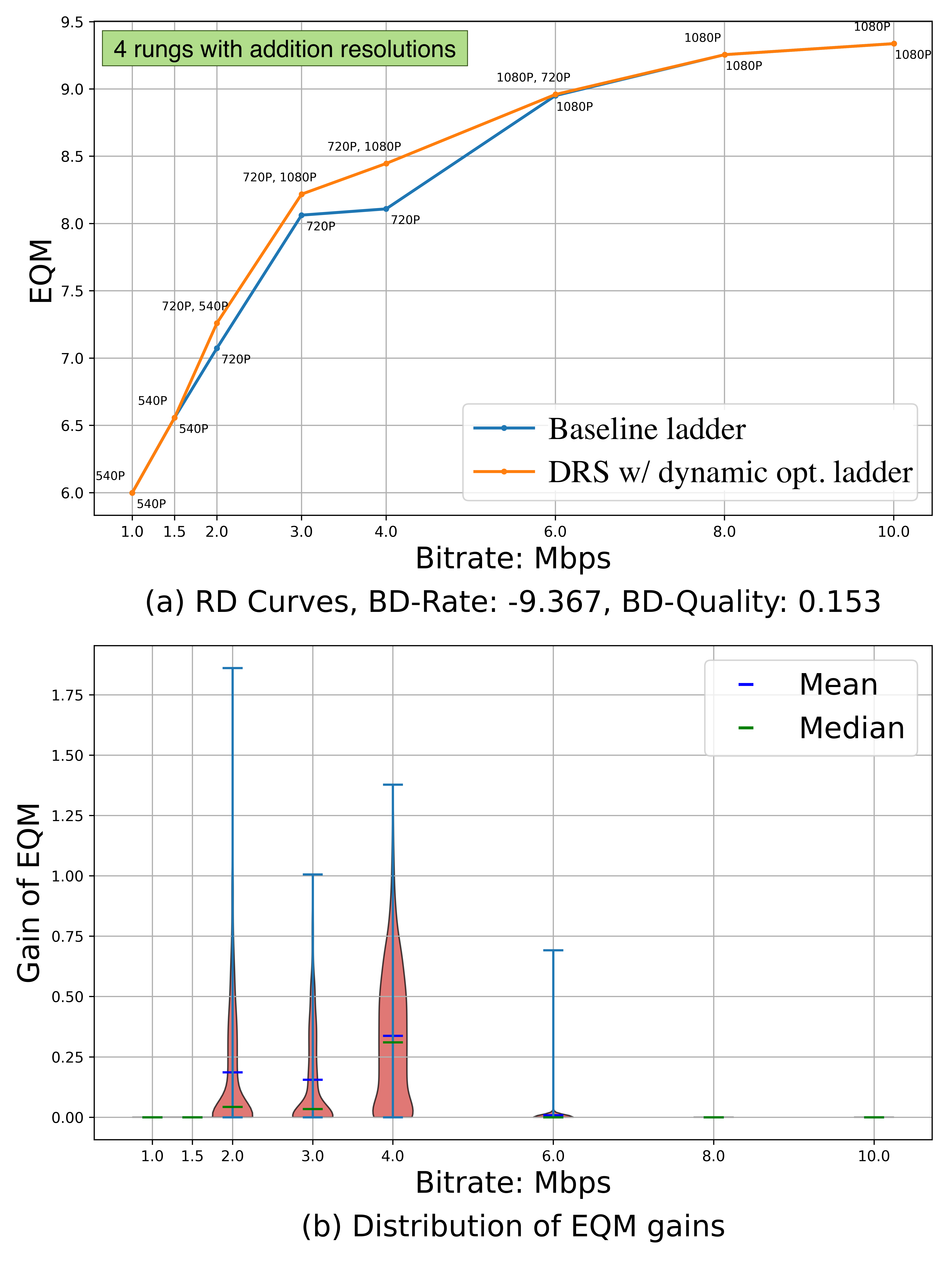}
\vspace{-3mm}
\caption{Performance evaluation of DRS using AVC-EQM scores on sports content. (a) RD curves for the baseline static ladder (blue) and DRS with the dynamic optimized ladder (orange). (b) Distribution of quality score gains across all GOPs; mean and median are marked by blue and green lines.
}
\vspace{-3mm}
\label{fig:TNF_EQM_Gain}

\end{figure}

% \begin{figure}[!t]
% \centering
% \centerline{\includegraphics[width=8.5cm]{Figures/AVC_POC/Best_resolution_prob_NBA_AVQM_NR.png}}

% \caption{Statistical analysis of resolution-quality relationships utilizing EQM scores for basketball video content.
% }
% \label{fig:NBA_AVQM_Prob}
% \end{figure}

% \begin{figure}[!t]
% \centering
% \centerline{\includegraphics[width=8.5cm]{Figures/AVC_POC/Gain_EQM_NBA_012225_HORNETS_GRIZZLIES_TEST_SDR_1_1_30min.mxf.png}}

% \caption{Performance evaluation for dynamic resolution switching using EQM scores for basketball video content. 
% }
% \label{fig:NBA_AVQM_Gain}
% \end{figure}

\subsection{AVC-EQM based dynamic resolution switching}
\label{prototype}
\autoref{tab:TNF_ladder} presents the optimized bitrate ladder for sports content based on AVC-EQM scores. We evaluate the content optimized ladder and the dynamic optimized augmented ladder on a 30-minute test sequence that is distinct from the one used for representation selection. \autoref{fig:TNF_EQM_Gain} (a) shows the rate-quality curves comparing the baseline against our proposed DRS pipeline with the dynamic optimized ladder, using AVC-EQM (on a 0--10 scale) as the VQM. The results demonstrate that the proposed DRS framework achieves a significant performance gain over the baseline static ladder.
\autoref{fig:TNF_EQM_Gain} (b) shows the distribution of AVC-EQM score improvements across GOPs, highlighting that some GOPs exhibit notable quality enhancement. Furthermore, substantial gains are observed particularly at 3 and 4\,Mbps, demonstrating the efficacy of our method in this critical resolution cross-over region.
While our evaluation focuses on sports content, the proposed DRS pipeline is designed to be generic: it selects the resolution with the highest VQM score at each bitrate, regardless of the content type. In addition, AVC-EQM and the ladder statistics can be re-estimated for other content using suitable subjective data, without modifying the DRS pipeline.
\autoref{tab:TNF_POC_Gain} reports the results on two 30-minute test sequences from different sports. On average, our method achieves a BD-rate saving of 8.97\% and a BD-quality gain of 0.152 (on a 10-point scale). 
These results, together with the minimal overhead of AVC-EQM, validate DRS as a practical and effective solution for QoE optimization in live streaming.

\begin{table}[!thbp]
\centering
\caption{Performance comparison between the content optimized ladder and our proposed DRS framework using an augmented ladder (4 additional representations) on two 30-minute sports sequences.}
\vspace{-2mm}
\label{tab:TNF_POC_Gain}
\begin{tabular}{ccccc}
\toprule
 & & Content Opt. & Dynamic Opt. \\ \midrule
\multirow{2}{*}{Sports 1} & BD-Rate    & -5.097  & \textbf{-9.367}\\
& BD-Quality & 0.082  & \textbf{0.153} \\ \midrule
\multirow{2}{*}{Sports 2} & BD-Rate    & -5.274  & \textbf{-8.567}\\
& BD-Quality & 0.091  & \textbf{0.151} \\ 
\bottomrule
\end{tabular}
\vspace{-3mm}
\end{table}

\section{Conclusion}
In this work, we proposed a dynamic resolution switching pipeline tailored for live streaming. The core of the method is selecting the optimal resolution, as measured by an efficient VQM, at any given bitrate during encoding. We first demonstrated the efficacy of the proposed AVC-EQM, validating its high correlation with subjective quality, its accuracy in estimating resolution cross-over points, and its runtime efficiency. We then extended the traditional static bitrate ladder by incorporating a statistical analysis of user-side bitrate distributions to maximize QoE, presenting a principled approach for selecting the additional representations. 
Empirical evaluation on real-world sports video sequences confirms significant quality improvements, demonstrating the overall effectiveness and practicality of the proposed method.
\vfill\pagebreak 
\bibliographystyle{IEEEbib}
\bibliography{strings,refs}

\end{document}